\documentclass[aps,prl]{revtex4}

\usepackage{graphicx}

\begin{document}

\title{Thomas rotation and Mocanu paradox -- not at all paradoxical}

\author{Z.~K.~Silagadze}

\affiliation{Budker Institute of Nuclear Physics and Novosibirsk State
University, 630 090, Novosibirsk, Russia}
\email{silagadze@inp.nsk.su}
%\date{\today}

\begin{abstract}
Non-commutativity of the Einstein velocity addition, in case of non-collinear
velocities, seemingly gives rise to a conflict with reciprocity principle.
However, Thomas rotation comes at a rescue and the paradox is avoided. It is
shown that such a resolution of the so called Mocanu paradox is completely 
natural from the point of view of basic premises of special relativity. 
\end{abstract}

\maketitle

\section{Introduction}
It should be clear, after a hundred years of development of special 
relativity, that to search a logical contradiction or paradoxes in it  
is the same as to search a logical inconsistency in non-euclidean
geometry (in fact, special relativity \emph{is} a kind of non-euclidean
geometry -- the Minkowski geometry of space-time). Surprisingly, however,
such efforts have never been abandoned. Some ``paradoxes'' are helpful
nevertheless because their resolution reveals the roots of our confusion
and, therefore, enhances our comprehension of special relativity.

The Mocanu paradox \cite{1,2,3} is an interesting paradox of this kind 
whose resolution makes clear some our misconceptions about space and time, 
deeply rooted in Newtonian intuition, which are notoriously hard to 
eliminate in physics students even after years of study of modern physics.

Although the resolution of this ``paradox'' is already available in the
literature (see \cite{3,4,5,6}), ``their arguments and mathematical formulas 
in terms of coordinates do not give an evident physical explanation of the 
paradox, though it became clear that the paradox was related somehow to the 
Thomas rotation'' \cite{6}.

It is the aim of this article to demonstrate by elementary means that there is 
nothing especially paradoxical about the Thomas rotation as far as it is 
considered with regard to the Mocanu paradox. To emphasize the physical
concepts involved, rather than mathematical formalism, we consider not the
most general case of the Mocanu paradox. However, the special case considered
already involves all necessary ingredients.  

\section{The Mocanu paradox}
Suppose a reference frame $S^\prime$ moves with the velocity $v$ with
respect to the frame ``at rest'', $S$, along its $x$-axis, and a frame
$S^{\prime\,\prime}$ moves with the velocity $v^\prime$  with respect to 
the frame $S^\prime$  along its $y^\prime$-axis. It is assumed that the
corresponding axes  of the frames $S$ and $S^\prime$ are parallel to each 
other, as do axes of the frames $S^\prime$ and $S^{\prime\,\prime}$. Then
the velocity $\vec{u}$ of $S^{\prime\,\prime}$ relative to $S$ is given
by the relativistic velocity addition law 
\begin{equation}
u_x=\frac{v^\prime_x+v}{1+\frac{v^\prime_x v}{c^2}}=v,\;\;
u_y=\frac{v^\prime_y}{\gamma\left (1+\frac{v^\prime_x v}{c^2}\right )}=
\frac{v^\prime}{\gamma},\;\;u_z=\frac{v^\prime_z}{\gamma\left (1+
\frac{v^\prime_x v}{c^2}\right )}=0,
\label{u-vel}
\end{equation}
where $\gamma$ is the Lorentz factor corresponding to the velocity $v$
$$\gamma=\frac{1}{\sqrt{1-(v/c)^2}}=\frac{1}{\sqrt{1-\beta^2}}.$$

According to the reciprocity principle \cite{7}, if $S^\prime$ moves relative
to $S$ with velocity $\vec{v}$, then $S$ moves relative to $S^\prime$ with
velocity $-\vec{v}$. Therefore, in the frame $S^{\prime\,\prime}$, the frame
$S^\prime$ moves along the $y^{\prime\,\prime}$ axis with the velocity 
$-v^\prime$, while in the frame $S^\prime$, the frame $S$ moves along the 
$x^\prime$ axis with the velocity $-v$. Compared to the previous situation, 
the roles of the $x$ and $y$ axes are interchanged, as are the roles
of $v$ and $v^\prime$ (with additional change of sign). Therefore, the 
velocity addition formula gives the velocity $\vec{u}^\prime$ of the frame
$S$ relative to $S^{\prime\,\prime}$
\begin{equation}
u^\prime_x=-\frac{v}{\gamma^{\,\prime}},\;\; u^\prime_y=-v^\prime,\;\;
u^\prime_z=0,
\label{uprim-vel}
\end{equation}  
where $\gamma^{\,\prime}$ corresponds to the velocity $v^\prime$.

Of course, it is possible to obtain all this by using the general formula 
for relativistic addition of non-collinear velocities \cite{3}
\begin{equation}
\vec{u}=\vec{v}\oplus\vec{v}^{\,\prime}=\frac{\vec{v}+\vec{v}^{\,\prime}}{
1+\frac{\vec{v}\cdot\vec{v}^{\,\prime}}{c^2}}+\frac{1}{c^2}\,
\frac{\gamma}{\gamma+1}\,\frac{\vec{v}\times
(\vec{v}\times\vec{v}^{\,\prime})}{1+\frac{\vec{v}\cdot\vec{v}^{\,\prime}}
{c^2}},
\label{vadd}
\end{equation}
from which a non-commutativity of this addition is clearly seen, but for our
purposes even simpler particular case of this formula for collinear 
velocities, (\ref{u-vel}), suffices if carefully used.

According to the reciprocity principle, the velocity of $S^{\prime\,\prime}$
relative to $S$ should be $-\vec{u}^\prime=\vec{v}^{\,\prime}\oplus\vec{v}$,
but it clearly does not equal to $\vec{u}=\vec{v}\oplus\vec{v}^{\,\prime}$.
And this constitutes the content of the Mocanu paradox: what is the correct
velocity of $S^{\prime\,\prime}$ relative to $S$, $\vec{v}\oplus\vec
{v^\prime}$ or $\vec{v}^{\,\prime}\oplus\vec{v}$, and how we can account for 
the reciprocity principle in this case?

We can discard a possibility that the reciprocity principle is violated from
the very beginning. In fact, it is possible and even preferable to base 
special relativity on this intuitively evident principle, instead of highly 
counter-intuitive second postulate (see \cite{8} and references therein).

\section{Resolution of the Mocanu paradox}
The key idea in resolution of the Mocanu paradox is the realization of 
the fact that space in special relativity is in fact more relative than 
space in the non-relativistic physics \cite{6}, although this can hardly be 
guessed by merely comparing the Galilean transformation $x^\prime=x-vt$, 
which describes relativity of space for non-relativistic observers, to its 
relativistic counterpart $x^\prime=\gamma(x-vt)$. In words of Minkowski, 
``space by itself, and time by itself are doomed to fade away into mere 
shadows, and only a kind of union of the two will preserve an independent 
reality'' \cite{9}.

The vectors  $\vec{v}\oplus\vec{v}^{\,\prime}$ and $\vec{v}^{\,\prime}\oplus
\vec{v}$ are defined in different reference frames $S$ and $S^{\prime\,
\prime}$ and, therefore, in different  spaces. It makes no sense to compare 
them unless the axes of $S$ and $S^{\prime\,\prime}$ are made parallel in 
some well defined way.

Axes of the $S$ and $S^\prime$, as well as axes of the  $S^\prime$ and 
$S^{\prime\,\prime}$ frames are assumed to be parallel, as mentioned above.
What conclusion we can draw then about the mutual orientation of the $S$ and 
$S^{\prime\,\prime}$ frames axes?

In the frame $S^\prime$, the $x^{\prime\,\prime}$ axis is given by the 
equation (we will drop $z$-coordinate as it is irrelevant in our planar case)
$$y^\prime=v^\prime t^\prime.$$
Then, according to Lorentz transformations, we conclude that in the frame $S$
the $x^{\prime\,\prime}$ axis is given by the equation
$$y=v^\prime \gamma \left (t-\frac{v}{c^2}\,x\right ).$$
Therefore, from the point of view of $S$, the $x^{\prime\,\prime}$ axis is
inclined clockwise relative to the $x$ axis by an angle $\alpha$ so that
\begin{equation}
\tan{\alpha}=\beta\beta^\prime\gamma.
\label{alpha}
\end{equation}
There is nothing paradoxical in this change of inclination. At least nothing
more paradoxical than the lack of absolute simultaneity from which it stems.
Note that such a change of inclination is used to resolve some pole-and-barn
type paradoxes \cite{10,11}.

\begin{figure}[htp]
\centering
\includegraphics[width=0.6\textwidth]{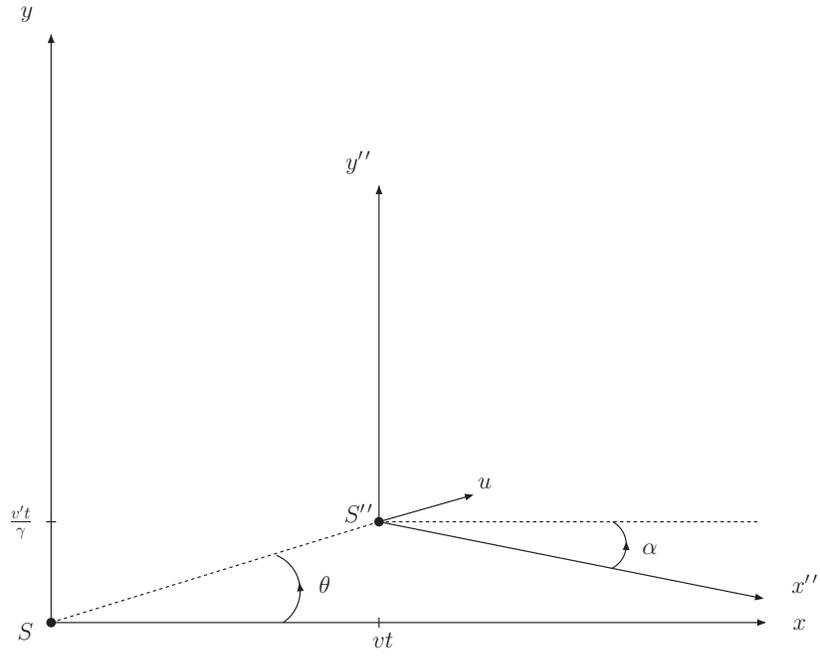}
\caption{Orientations of the $S^{\prime\,\prime}$ axes as perceived in the 
frame $S$.}
\label{fig1} 
\end{figure}

Analogously, $y^{\prime\,\prime}$ axis is given in the frame $S^\prime$ by
the equation $x^\prime=0$, which in the frame $S$ transforms into
$$\gamma(x-vt)=0.$$
Therefore, $y^{\prime\,\prime}$ axis is given in the frame $S$ by the 
equation $x=vt$ and, consequently, remains parallel to the $y$ axis.  
Fig.\ref{fig1} summarizes the orientations of the  $x^{\prime\,\prime}$ and
$y^{\prime\,\prime}$ axes as seen by an observer in the $S$ reference frame.

We need some refinement here. Because of the finite speed of light, we should
distinguish between what Rindler calls \cite{12} world-picture and world-map.
World-picture is what an observer actually sees at any given moment of time, 
a snapshot which records distant objects at different moments of the past. 
World-map, on the contrary, is the set of events that the observer considers
to have occurred in the world at that instant of time. Special relativity
operates with world-maps, Lorentz transformation being an instrument which 
relates two world-maps of different inertial frames. Therefore, when we speak 
rather loosely about what an observer sees or perceives, actually we have 
in mind the world-map of this observer. With this caveat, let us continue 
and find how the situation described by Fig.\ref{fig1} is transformed in 
the frame $S^{\prime\,\prime}$. 

\begin{figure}[htp]
\centering
\includegraphics[width=0.6\textwidth]{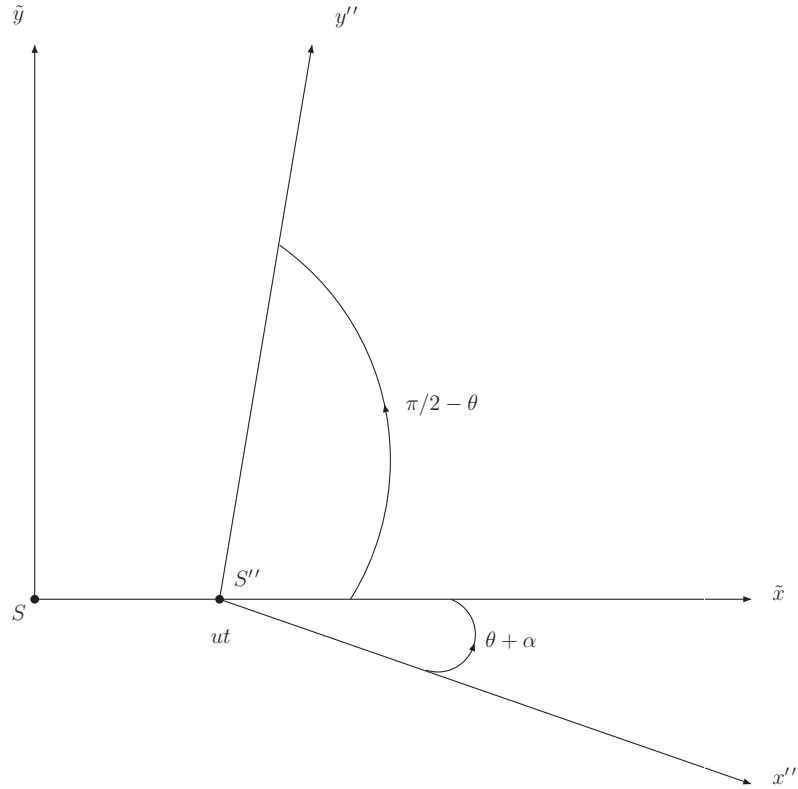}
\caption{Orientations of the $S^{\prime\,\prime}$ axes as perceived in the 
frame $S$ relative to the $\tilde x$ and $\tilde y$ axes.}
\label{fig2} 
\end{figure}

First of all, let us introduce another set of axes $\tilde x,\tilde y$ and
$\tilde x^{\prime\,\prime},\tilde y^{\prime\,\prime}$, so that $\tilde x$
and $\tilde x^{\prime\,\prime}$ are parallel to $\vec{u}$ and, therefore, $S$
and $S^{\prime\,\prime}$ equipped with these axes are in a standard 
configuration. In these new axes, Fig.\ref{fig1} is changed into  
Fig.\ref{fig2}.

\begin{figure}[htp]
\centering
\includegraphics[width=0.4\textwidth]{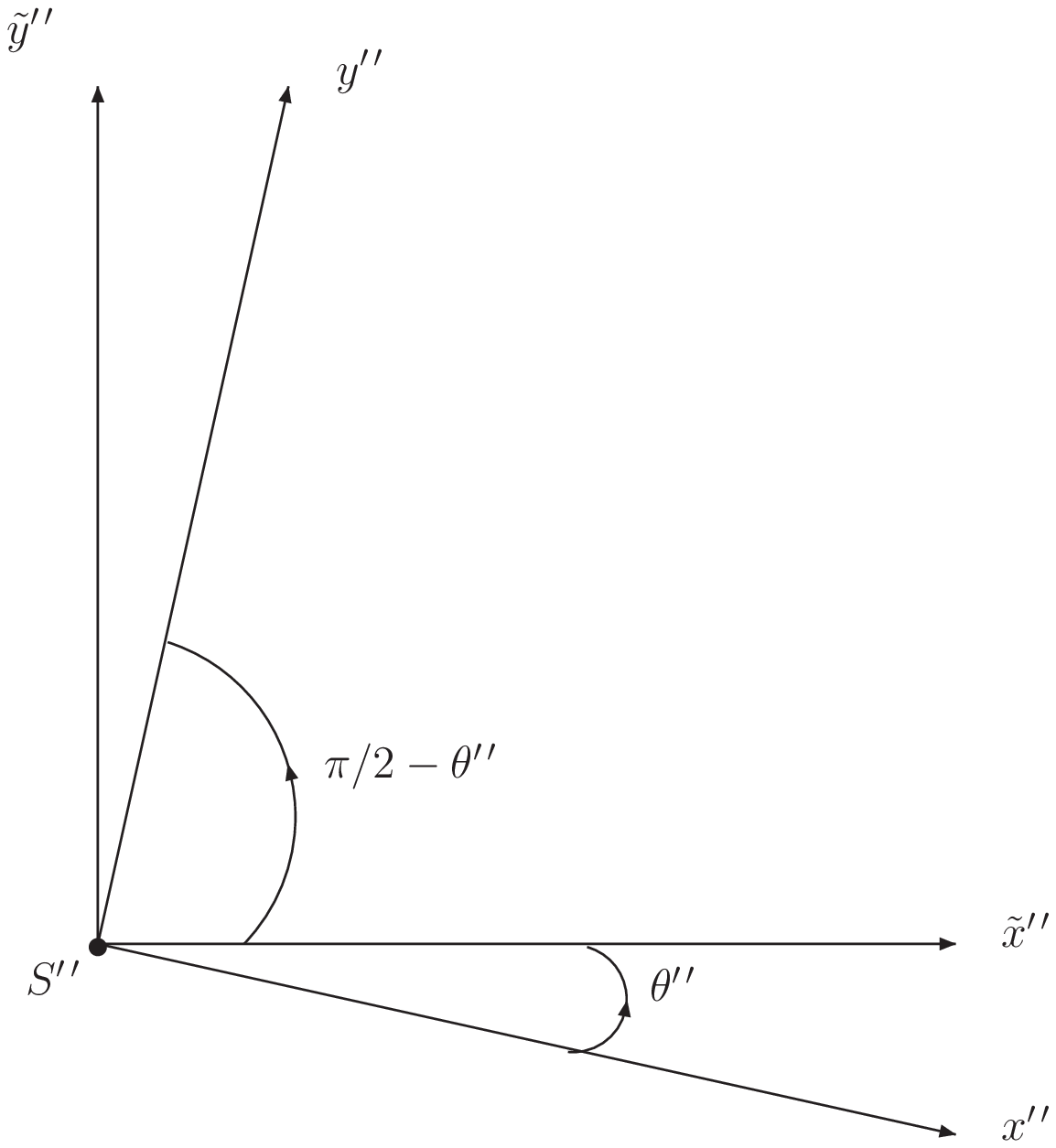}
\caption{Orientations of the $S^{\prime\,\prime}$ axes $x^{\prime\,\prime}$ 
and $y^{\prime\,\prime}$ as perceived in the 
frame $S^{\prime\,\prime}$ relative to $\tilde x^{\prime\,\prime}$ 
and $\tilde y^{\prime\,\prime}$.}
\label{fig3} 
\end{figure}

But
$$\tan{(\theta+\alpha)}=\frac{\tan{\theta}+\tan{\alpha}}{1-\tan{\theta}\,
\tan{\alpha}}=\frac{\frac{\beta^\prime}{\beta\gamma}+\beta\beta^\prime\gamma}
{1-\beta^{\prime\,2}}=\frac{\gamma\beta^\prime\gamma^{\,\prime\,2}}{\beta},$$
and
$$\tan{(\frac{\pi}{2}-\theta)}=\cot{\theta}=\frac{\beta\gamma}
{\beta^\prime}.$$
Therefore, the equation which defines  $x^{\prime\,\prime}$ axis in the 
frame $S$ looks like
\begin{equation}
\tilde y=-\frac{\gamma\beta^\prime\gamma^{\,\prime\,2}}{\beta}(\tilde x-ut),
\label{eqs}
\end{equation}
while the equation for the  $y^{\prime\,\prime}$ axis is
\begin{equation}
\tilde y=\frac{\beta\gamma}{\beta^\prime}(\tilde x-ut).
\label{eqss}
\end{equation}
Let us apply now the Lorentz transformation
$$\tilde x=\gamma_u(\tilde x^{\prime\,\prime}+ut^{\prime\,\prime}),\;\;
t=\gamma_u(t^{\prime\,\prime}+\frac{u}{c^2}\tilde x^{\prime\,\prime}),\;\;
\tilde y=\tilde y^{\prime\,\prime},$$
to change world-map from $S$ to $S^{\prime\,\prime}$. As a result, we get 
from (\ref{eqs})
\begin{equation}
\tilde y^{\prime\,\prime}=-\frac{\gamma\beta^\prime\gamma^{\,\prime\,2}}
{\beta}\gamma_u(1-\beta_u^2)\tilde x^{\prime\,\prime}=-\frac{\gamma
\beta^\prime\gamma^{\,\prime\,2}}{\beta\gamma_u}\tilde x^{\prime\,\prime}=
-\frac{\beta^\prime\gamma^{\,\prime}}{\beta}\tilde x^{\prime\,\prime},
\label{eqI}
\end{equation}
where at the last step we have used
$$\gamma_u=\gamma_{\vec{v}\oplus\vec{v}^{\,\prime}}=\gamma\gamma^{\,\prime}
\left (1+\frac{\vec{v}\cdot\vec{v}^{\,\prime}}{c^2}\right )=\gamma
\gamma^{\,\prime}.$$
Analogously, (\ref{eqss}) transforms into
\begin{equation}
\tilde y^{\prime\,\prime}= \frac{\beta\gamma}{\beta^\prime\gamma_u}
\tilde x^{\prime\,\prime}=\frac{\beta}{\beta^\prime\gamma^{\,\prime}}
\tilde x^{\prime\,\prime}.
\label{eqII}
\end{equation}
Equations (\ref{eqI}) and (\ref{eqII}) show that, from the point of view of 
an observer in the frame $S^{\prime\,\prime}$, $x^{\prime\,\prime}$ and 
$y^{\prime\,\prime}$ axes are inclined with respect to the $\tilde x^{\prime\,
\prime}$ axis (and, hence, with respect to the line of relative motion) with
angles $-\theta^{\prime\,\prime}$ and $\pi/2-\theta^{\prime\,\prime}$
respectively, as shown in Fig.\ref{fig3}.

At that
\begin{equation}
\tan{\theta^{\prime\,\prime}}=\frac{\beta^\prime\gamma^{\,\prime}}{\beta}.
\label{thetapp}
\end{equation}

While naively one should expect the inclination angle $\theta^{\prime\,
\prime}$ to be the same angle $\theta$ by which the $x$ axis is inclined 
with respect to the $\tilde x$ axis in the frame $S$, as it would be in 
case of parallel  $x$ and  $x^{\prime\,\prime}$ axes, it is not, because
\begin{equation}
\tan{\theta}=\frac{\beta^\prime}{\beta\gamma}.
\label{theta}
\end{equation}

As we see, although axes of the frames $S$ and $S^\prime$, as well as
$S^\prime$ and $S^{\prime\,\prime}$, were rendered  parallel, the axes
of the frames $S$ and $S^{\prime\,\prime}$ turned out not to be parallel
in any meaningful way. Space for relativistic observers are more relative
than for non-relativistic observers and we should be very careful while
interpreting the results of several consecutive non-collinear boosts.

The difference $\epsilon=\theta^{\prime\,\prime}-\theta$ is the notorious
Thomas rotation and it provides a ready explanation of the Mocanu paradox:
an observer in the frame $S^{\prime\,\prime}$ really perceives $-\vec{u}$
as the velocity of the frame $S$, in agreement with the reciprocity 
principle, but projects this vector of relative velocity onto 
$x^{\prime\,\prime}$ and $y^{\prime\,\prime}$ axes to get its components
$$u^\prime_x=-u\cos{\theta^{\prime\,\prime}}=-\frac{u}{\sqrt{1+\tan^2
{\theta^{\prime\,\prime}}}}=-\frac{v}{\gamma^{\,\prime}},\;\;
u^\prime_y=u^\prime_x\tan{\theta^{\prime\,\prime}}=-v^\prime.$$

As we see, Thomas rotation reconciles the reciprocity principle with the 
non-commutativity of relativistic velocity addition.

\section{Concluding remarks}
Thomas rotation and Thomas precession are often considered as obscure 
relativistic effects which have generated a huge, sometimes confusing 
literature \cite{13}. Nevertheless, this phenomena ``can be quite naturally 
introduced and investigated in the context of a typical introductory course 
on special relativity, in a way that is appropriate for, and completely 
accessible to, undergraduate students'' \cite{14}. I think the Mocanu 
paradox provides a very useful possibilities in this respect.

The resolution of the paradox presented in this article was essentially 
given by Ungar \cite{3} years ago. I hope, however, that the above 
presentation is simpler and clarifies some confusion. For example, it is 
claimed in \cite{3} that an observer in $S$ sees the axes of 
$S^{\prime\,\prime}$ rotated relative to his own axes by a Thomas rotation 
angle, $\epsilon$. However, this is not correct. The observer in  $S$ 
``sees'' what is depicted in Fig.\ref{fig1}. Thomas rotation angle, 
$\epsilon$, emerges when we compare the orientation of $S^{\prime\,\prime}$ 
axes, as actually seen by an observer in $S^{\prime\,\prime}$, to the naive 
expectation of the observer in $S$ what the observer in $S^{\prime\,\prime}$ 
should see if the transitivity of parallelism is assumed between different 
inertial reference frames.  

Thomas rotation is very basic phenomenon in special relativity which follows
quite naturally from its basic premises, as was demonstrated above. It is as 
basic as the time dilation and length contraction and is no more paradoxical 
than these well known effects of special relativity. Of course, this does 
not mean that it is trivial. It took years before ``evidence that Einstein's 
addition is regulated by the Thomas precession has come to light, turning 
the notorious Thomas precession, previously considered the ugly duckling 
of special relativity theory, into the beautiful swan of gyrogroup and 
gyrovector space theory'' \cite{15}. At this more advanced level, you can 
enjoy also other non-Euclidean facets of relativistic velocity space 
\cite{16}, from which the geometrical meaning of Thomas rotation, first 
discovered  by the famous French mathematician \'{E}mile Borel long before 
Thomas found the precession effect \cite{17}, becomes evident.

\end{document}